\documentclass[11pt,letterpaper]{article}
\usepackage[final]{graphics}
\usepackage{graphicx}
\usepackage{makeidx}
\usepackage{authblk} 
\usepackage{epstopdf}
\usepackage{listings}
\usepackage{pstricks,tikz}
\usepackage[absolute,overlay]{textpos}
\usepackage{appendix}

\usepackage{amsmath, amssymb}
\usepackage{wrapfig}
\usepackage{color,bm}
\usepackage{natbib}
\usepackage[utf8]{inputenc}
\usepackage[portrait]{geometry}
\usepackage{setspace}
\usepackage{csquotes}

\graphicspath{{figures/}}

\begin{document}

\title{Characterizing Interface Topology in Multiphase Flows using Skeletons}

\author[1]{Xianyang Chen}
\author[1]{Jiacai Lu}
\author[2,3]{St\'ephane Zaleski}
\author[1]{Gr\'etar Tryggvason}

\affil[1]{Johns Hopkins University, Baltimore, MD, USA}
\affil[2]{Sorbonne Universit\'e et CNRS, Institut Jean Le Rond $\partial$’Alembert, UMR, Paris 7190, France}
\affil[3]{Institut Universitaire de France, Paris, France}

\date{\today}

\maketitle

\begin{abstract}
The unsteady motion of a gas-liquid interface, such as during splashing or atomization, often results in complex liquid structures embedded in the ambient fluid. Here we explore the use of skeletonization to identify the minimum amount of information needed to describe their geometry. We skeletonize a periodic liquid jet by a modification of a recently introduced approach to coarsen multiphase flows while retaining a sharp interface. The process consists of diffusing an index function and at the same time moving the interfaces with it, until they ``collapse'' into each other and form skeletons. The skeleton represents the basic topology of the jet and we also keep track of how much the interface is moved (or how much volume is ``accumulated'') during the process, which can be used to approximately reconstruct the jet. We explore various quantitative measures to characterize and distinguish the skeletons. Those include standard morphometrics such as branch length distribution, after segmenting the skeletons into branches, and a more sophisticated representation of the skeleton structures called Topology Morphology Descriptor (TMD), to obtain an ``equivalent'' description of the skeletons by retaining information about the topology in a compact way.
\end{abstract}

\section{Introduction}

Using singular structures as approximations for finite-size objects is common in fluid mechanics. Point vortices represent smooth finite-size vortices, infinitely thin boundary layers replace finite-thickness ones and in multiphase flows point particles are commonly used to model finite-size drops, bubbles and solid particles. Likewise, thin films are often modeled as one or two-dimensional lines and sheets. When the shape of the object that is being singularized is simple enough, identifying the singular structure is straightforward (such as for nearly spherical particles or nearly flat films), but extending singularization to complex structures requires a more formal approach. Here, we explore how to develop singular approximations to complex flow objects such as jets that are breaking up, and where the shape of the singular structure may not be easily identified, using skeletonization. We are, in particular, concerned with flow objects that may have non-trivial topology, such as branches, holes and isolated segments.  

Skeletonization, where the dimension of a complex two or three-dimensional object is reduced by one or two dimensions by contracting it in the appropriate way, is used for compact representation of image objects in a variety of applications such as medical imaging, computer graphics, visualization and shape analysis. Line-skeletons are, for example, used to create a collision-free path through a 3D object for virtual navigation and virtual endoscopy (\cite{he2001reliable,wan2001distance}). In traditional computer graphics, objects are represented by stick-like figures (IK-skeletons) and used to facilitate animations (\cite{bloomenthal2002medial}). Another common application of skeletons is registration, which helps to align two images taken with different modalities (MRI, CT) from the same patient (\cite{pizer2003multiscale}). In fluid dynamics extracting the lines (skeletons) from the vortex cores has been used by several authors (\cite{banks1994vortex,peikert1999parallel,roth1998higher,linnick2005vortex,bader2019extraction}) for better flow visualization and analysis. Various techniques have been introduced to obtain skeletons, depending on the data that is available (volumetric description by pixels or parameterized surfaces, for example) and the intended applications. Skeletonization algorithms can be categorized into four main classes according to \cite{cornea2007curve}: (1) thinning and boundary propagation; (2) distance field based; (3) geometric; and (4) general-field functions. In many methods concepts from more than one class are used to produce a skeleton. Based on this categorization, here we combine the surface thinning with a general-field function, and gradually shrink the surface by moving it along with the diffused field. The most similar approach that we can find in the literature is \cite{schirmacher1998boundary}, where the skeleton is obtained by shrinking the surface in the direction indicated by the distance field. As in the original grassfire algorithm of \cite{Blum1967} our strategy is based on evolving the boundary in (pseudo) time, but we find the normal velocity by solving a parabolic diffusion equation instead of taking it to be a constant. For surveys of the various approaches, a discussion of the many challenges, and examples of applications, see \cite{cornea2007curve,saha2016survey,tagliasacchi20163d,saha2017skeletonization}, for example.


Skeletonization is a compact way of representing the topology, but we are also interested in how to distinguish skeletons from each other, with minimum level of measurements. This is usually referred to branching morphology or branching network, and used to analyze the statistical properties of many different systems that exhibit tree structures such as gorgonian corals (\cite{sanchez2003similar,cadena2010linking,brazeau1988inter}), cells or neurons (\cite{den2006branching,gillette2015topological1,gillette2015topological2,kim2012geometric}) and trees and river networks (\cite{pelletier2000shapes,dodds2000geometry,zhang2009quantization}). The standard measurements, including branch length, bifurcation ratio, number of branches and so on, are commonly used to characterize the branching structure. In biology and river systems, many branching structures appear to be self-similar across wide range of scales, and those are often studied using fractal dimension or Tokunaga parameters (\cite{masters2004fractal,turcotte1998networks,dodds2000scaling,zanardo2013american}). \cite{kanari2018topological} proposed a more sophisticated measurement of branching structures--Topology Morphology Descriptor (TMD) which retains enough topological information to allow systematic comparison between branching morphologies and has been used to classify different types of neocortical pyramidal cells (\cite{kanari2019objective}). Here we use a slightly modified version of the TMD algorithm to study fluid skeletons. 

For multiphase flows, we believe skeletonization can be useful for at least three tasks
\begin{itemize}
\item Analyze complex flow structures by systematically eliminating scales and exposing the underlying topological structure.
\item Find the minimum information needed to describe the structures, and to construct an equivalent---in some sense---skeleton.
\item Build reduced order models, similar to point particles, but for more complex objects.
\end{itemize}
Here we will focus on the first two. We start by developing a strategy to construct the skeleton and examine how well it represents the original structure, and then we quantify the shape of the skeleton using a variety of quantitative measures describing both its spatial layout and its topology. Finally, we examine the use of TMD to represent the structures in a more compact way. We work  mostly with two-dimensional flows for simplicity, but show one example of the skeletonization of fully three-dimensional jet. 

\section{Method}

Reducing the dimensions of a complex structure can be done in may ways, as discussed in the references cited in the introduction. Here we use a modified version of a filtering technique introduced in \cite{chen2021interface}, where a diffusion equation is solved for a phase indicator function and the contour originally identifying the interface is tracked. 
We start by defining an indicator function that takes a different value in the different fluids
\begin{equation}
\chi ({\bm x}) = \left\{ 
  \begin{array}{l l}
\text{0 in fluid 0;}\\
\text{1 in fluid 1,}
  \end{array} \right.
\end{equation}
and smooth it slightly to give a continuous transition from one fluid to the other. The interface is identified by the $\chi_0=0.5$ contour. To simplify the interface we diffuse the indicator function by solving
\begin{equation}
{\partial \chi \over \partial \tau} =D\nabla^2 \chi.
\end{equation}
Notice that we can set the diffusion coefficient to unity ($D=1$) since a different value simply rescales $\tau$.
To move the interface with a specific contour, $\chi = \chi_s$ that can be different from the interface contour, we compute the ``diffusion velocity," 
\begin{equation}
{\bm u}_I=u_n {\bm n} \approx  -{ (\chi_s - \chi^* )  \over \vert \nabla \chi  \vert^2  \Delta  \tau }  \nabla \chi.
\label{interfacevelocity1}
\end{equation}
where $\chi^*$ is the old value of the index function at the interface. See \cite{chen2021interface} for details.

If the indicator function takes a constant value in each phase, then tracking the contour identified with the average value $\chi_0$ conserves the volume of both phases (within numerical errors), but by selecting a different contour value $\chi_s$ that is different from $\chi_0$ we move the interface into the phase identified by the value  of the indicator function that is closer to the selected contour value. To skeletonize fluid 0 we pick $0.1$ and for fluid 1 we select $0.9$. Picking different values only affects the total diffusion time (computational time) of the skeletonization process and has no impact on the structure of the final skeletons. 

The motion of each interface segment is stopped when it runs into another interface and the stationary interface forms the skeleton. The different parts of the interface stop moving at different times (or iterations) depending  on how far they move before encountering another interface. We keep track of the stopping time, $\tau_s$, for each point on the skeleton. As we will see in the next section the stopping time is closely correlated to the width of the original film, thus allowing us to assign mass to each interface point. Since we track the Lagrangian points on the interface throughout the process, the one-to-one correspondence of the points on the skeletons with the original surface is retained. This is beneficial when segmenting different parts of the original interface based on the skeleton segments for use in animation (\cite{cornea2007curve,sharf2007fly}). During the skeletonization process, the unstructured mesh on the interface is refined every pseudo time step and sometimes refining the mesh along the contour is also necessary. As will be seen later in section \ref{jetsection}, the performance of skeletonization also depends on the Eulerian grid resolution. In principle, being weakly sensitive to the boundary noise is inherently built into our skeletonization algorithm. This is referred to as skeleton ``robustness'' by \cite{cornea2007curve}. Other skeletonization methods, such as using a distance field to extract a medial axis, needs an additional operation called ``pruning'' to eliminate the noise in the skeletons (\cite{sharf2007fly}). In our method, we can also control the sensitivity of the skeletons to the interface noise by changing the constant diffusion coefficient into a nonuniform variable that is dependent on curvature $\kappa$. Details are shown in Appendix \ref{appendix}.

\section{Results}
\subsection{Demonstration of Skeletonization}
We start by applying the skeletonization algorithm to a simple 2D jet. The jet is obtained by solving the Navier-Stokes equations by a Front-tracking/Finite-volume solver on a $256\times256$ grid, with second order accuracy in space and time. The computational domain is a $1\times1$ square box, with periodic boundary condition in the $x$ direction and a full-slip wall boundary condition in the $y$ direction. The diameter of the jet is $d=0.3$. The jet has density $\rho_j=2.5$ and viscosity $\mu_j=0.01$ and the surrounding fluid has density $\rho_f=1.25$ and viscosity $\mu_f=0.001$. The density ratio is therefore $r=\rho_j/\rho_f=2$ and the viscosity ratio is $m=\mu_j/\mu_f=10$. The surface tension is $\sigma=0.005$. The initial velocity of the jet is $u_j=2$ and the ambient fluid is stationary so the velocity jump is $\Delta u=2$. These parameters give $Re=\rho_j \Delta u d/\mu_j=150$ and $We=\rho_j\Delta u^2d/\sigma=600$. The computational setup is similar to \cite{afanador2021effect}.

\begin{figure}
    \centering{\includegraphics[scale=0.3]{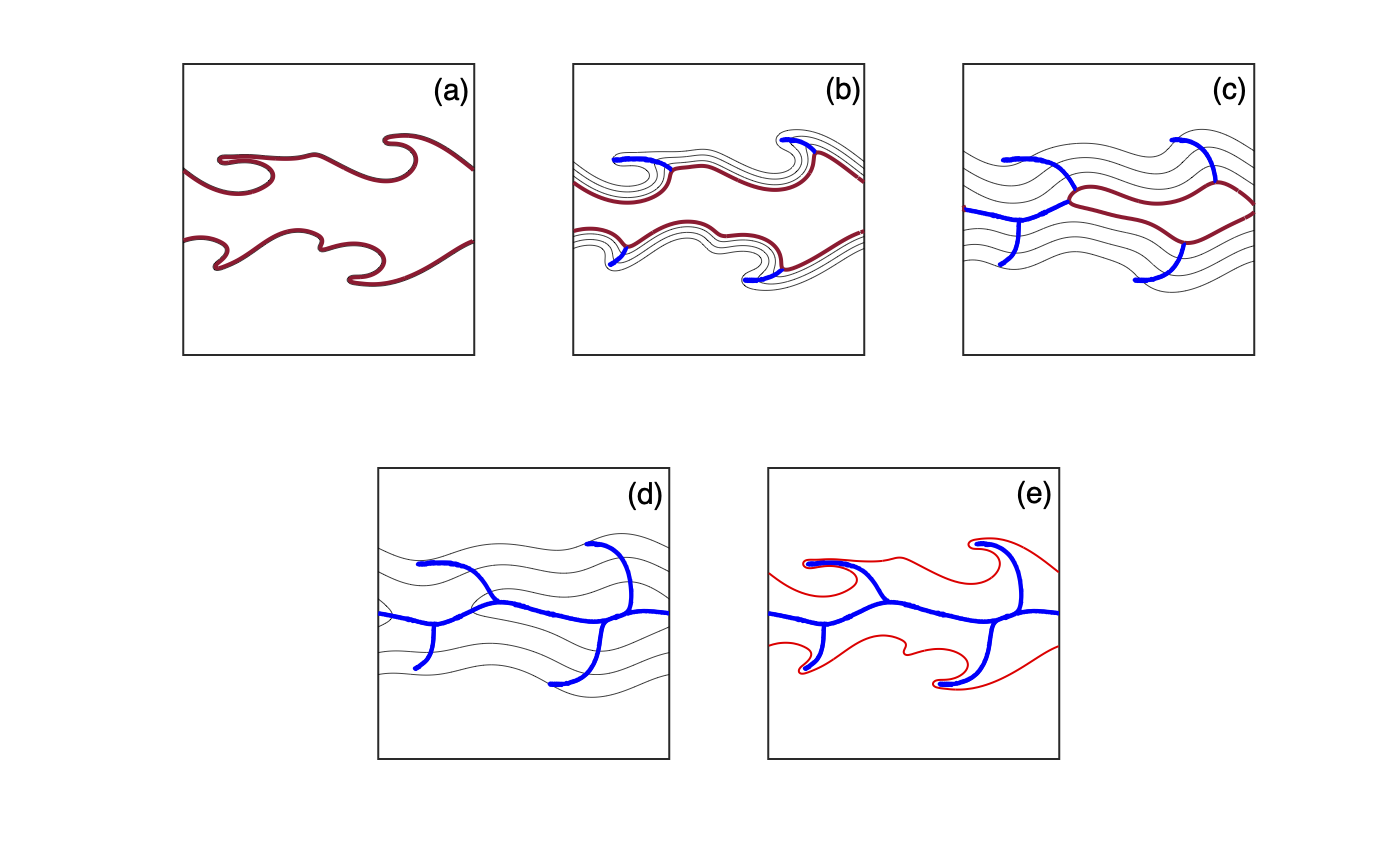} }
    \caption{(a)-(d): The skeletonization of the jet at different pseudo time $\tau $ (a) $\tau=0$, (b) $\tau=4\times10^{-4} $, (c) $\tau=2.5\times10^{-3} $ and (d) $\tau=5\times10^{-3} $. The black lines are the contour lines of the diffused indicator function $\Tilde{\chi} $. The interfaces that are close enough and collapsed into the skeletons are shown in blue and other parts where they still follow the contours are shown in red. (\textbf{e}) The final skeleton (blue) plot along with the original interface (red). }
    \label{evolution}
\end{figure}

Figure \ref{evolution} shows the evolution as the jet is skeletonized by integrating in pseudo time at computational time $t=0.43$, where \ref{evolution}(a) plots the original interface and \ref{evolution}(d) shows the skeleton. The Lagrangian points on the interface are moved with the diffused indicator function and forced not to move if skeletons are formed. The relatively small structures collapse rapidly into skeletons since the indicator function is diffused out quickly, according to figure \ref{evolution}(b), while for the large structures (plotted in red) take more time to collapse into skeletons. Once the whole jet has collapsed into thin films, the skeletonization process is done. For each part of the skeleton, we record the local pseudo time $\tau_s$ when the interface stops moving and the segment has skeletonized. As can be seen from figure \ref{evolution}(e), the skeleton reflects the intrinsic topology reasonably well, and in particular, short branches are not shortened. Moreover, the skeletonization algorithm automatically picks the obvious structures to form the skeletons and ignores the nonobvious ones (for example the small bump in \ref{evolution}(e)).

\subsection{Skeletonization of a Larger Jet} \label{jetsection}

To analyze the skeletonization in more detail, we apply the skeletonization algorithm to a larger jet. The computational domain is enlarged to an $8\times2$ rectangular box and solved on a $1024\times256$ grid. The density and viscosity for both phases as well as the velocity jump $\Delta u$ are the same as before. The diameter of the jet is set to $d=0.6$ and surface tension is doubled to $\sigma=0.01$, resulting in $Re=\rho_j \Delta u d/\mu_j=300$ and $We=\rho_j\Delta u^2d/\sigma=600$. The initial jet interface is perturbed slightly by adding 30 waves with random amplitude and wavenumbers, making sure that it includes the most unstable wavenumber, according to Kelvin-Helmholtz instability analysis. 

\begin{figure}
    \centering{\includegraphics[scale=0.4]{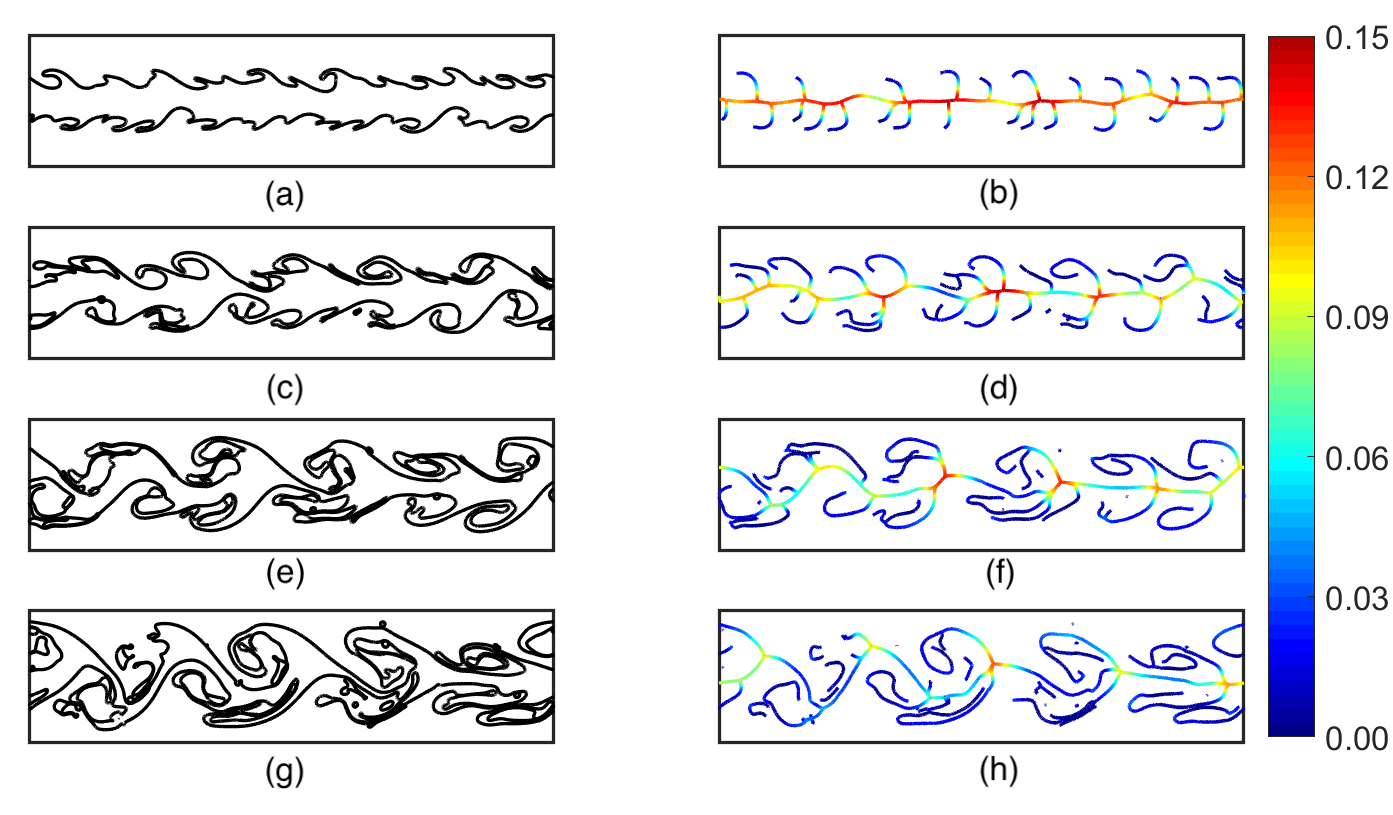} }
    \caption{The original interfaces (left column) and their skeletons (right column) at (a,b) $t=0.625$, (c,d) $t=1.25 $, (e,f) $t=1.875 $ and (g,h) $t=2.5 $. The skeletons are colored based on the local length scale $L_s$.}
    \label{colorful2dskeletons}
\end{figure}

Figure \ref{colorful2dskeletons} shows the jet and its skeleton at four different times. The skeletons are colored based on the local length scale $L_s$, which is related to the local collapsing pseudo time by $L_s=\sqrt{D\tau_s}$. Note that the grid resolution for the skeletonization is doubled ($2048\times512$) for the last 2 rows in order to get better quality skeletons, by reconstructing the indicator function on the finer grid using the current interface. Some spurious branches are formed even by doubling the grid resolution, which usually occur when thin films exist in the other phase. In principle, this can be solved by using fine enough grid resolution, although we have not done that due to the limitation of computational time. At the early stage the jet structure is relatively simple, with short branches connecting to the main trunk, resembling a fish-bone structure. At later time the branches start to grow longer, generate new branches, or detach from the main trunk and form branches not connected to the main contiguous tree. At the final stage the skeleton is longer and consists of a larger number of branches and isolated segments. In contrast to the original interfaces, where the topology becomes more complicated with time and harder to visualize, the skeleton exposes the underlying topology and provides a compact representation. 

\begin{figure}
    \centering{\includegraphics[scale=0.45]{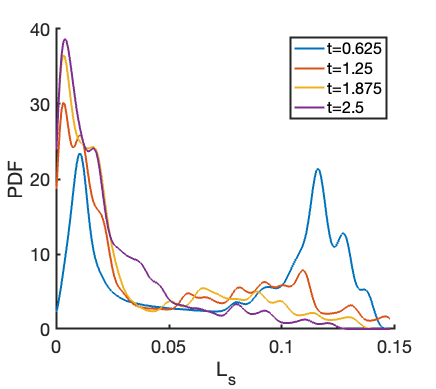} }
    \caption{The probability density function (pdf) of the length scale $L_s$ at 4 different times. }
    \label{pdflength}
\end{figure}

The local length scales $L_s$ are decreasing with time, as can be seen from the colors in the plot. The probability density distribution (pdf) of the local length scales for different times is plotted in figure \ref{pdflength}. At the beginning there are two peaks in the pdf of $L_s$, one for the large scales and the other for the small scales and both are about as frequent. At later time the small structures become more dominant and the large structures less so, indicating a gradual shift from large to small scales. 

\begin{figure}
    \centering{\includegraphics[scale=0.5]{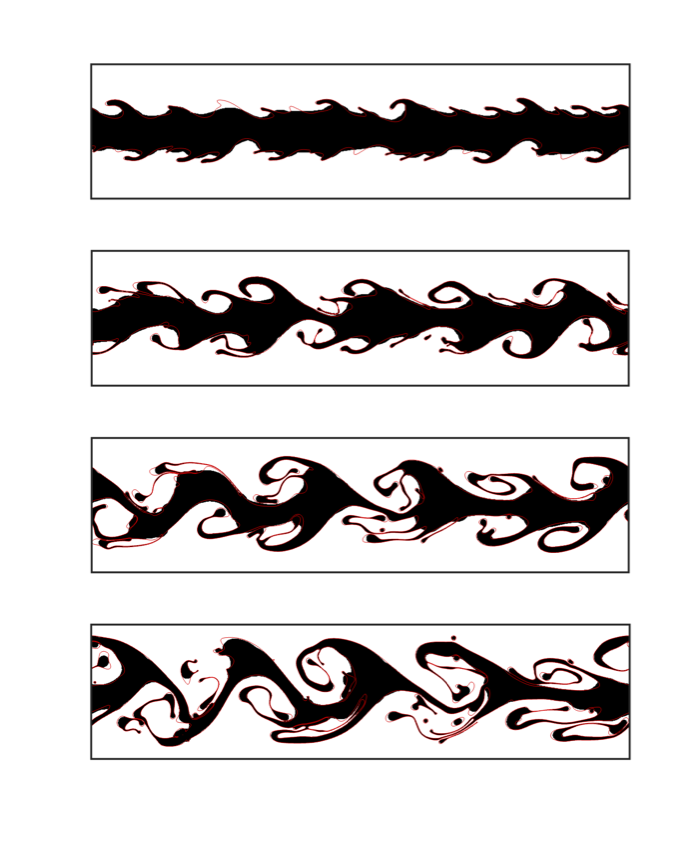} }
    \caption{The reconstructed indicator function field based on the skeletons are shown in black and the original interfaces in red, at the same times as in figure \ref{colorful2dskeletons}. }
    \label{reconinterface}
\end{figure}

\begin{figure}
    \centering{\includegraphics[scale=0.5]{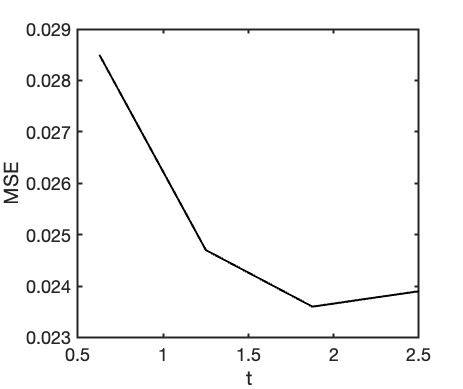} }
    \caption{The MSE for the reconstructed indicator fields versus the original ones versus time, with 4 data points corresponding to the times in figure \ref{reconinterface}. }
    \label{reconmse}
\end{figure}

One of the measures of the skeleton quality is the ability to reconstruct the original interface topology from the skeleton \citep{cornea2007curve}. We should not expect the original interface to be reconstructed exactly since the skeletons are supposed to eliminate small structures, although in principle, we can find the medial axis (surface) (\cite{amenta2001power, dey2004approximating,dey2006defining,katz2003untangling}) and reconstruct the interface exactly. The indicator function is reconstructed by computing the union of inscribed balls centered at every skeleton point, and simply setting the values to 1 inside the union of balls and 0 anywhere else. The radius of the balls is proportional to the skeleton length scale $L_s$. We determined the relationship between the radius of the balls and the length scale $L_s$ by numerical experiment, where a flat thin film is collapsed into a line skeleton, and found that $R_s=2.352L_s$. The reconstructed indicator function is plotted in figure \ref{reconinterface}, where the original interface is plotted in red. The reconstructed shape is very similar to the original jet, although some discrepancies can be seen, as expected. This further justifies the quality of skeletons. As can be seen from figure \ref{reconinterface}, structures with high aspect ratio (thin films) and with aspect ratio of 1 (circular drops) are reconstructed more accurately than structures with moderate aspect ratio (wavy or elliptical structures). Figure \ref{reconmse} shows the mean square error (MSE) of the reconstructed indicator fields with the original ones versus time. The $MSE$ is defiend as $MSE=\frac{1}{N}\sum_{j}^{N}(I_j-I^{recon}_j)^2$, where $I$ and $I^{recon}$ are the original and reconstructed indicator function and $N$ is the total number of grid points. The MSE is decreasing with time because at later time there are more thin films and drops while at the initial stage the wavy structures with moderate aspect ratio dominates and are not captured by the skeletonization. 

\subsection{Quantitative Measures of Skeletons}

A visual inspection of figure \ref{colorful2dskeletons} suggests that the skeleton does represent the original jet reasonably well and using the skeleton as a compact representation of the flow structures is likely to be an important application. In this section, we further examine the topology by computing various quantities describing the skeleton. 

\begin{figure}
    \centering
    \includegraphics[scale=0.4]{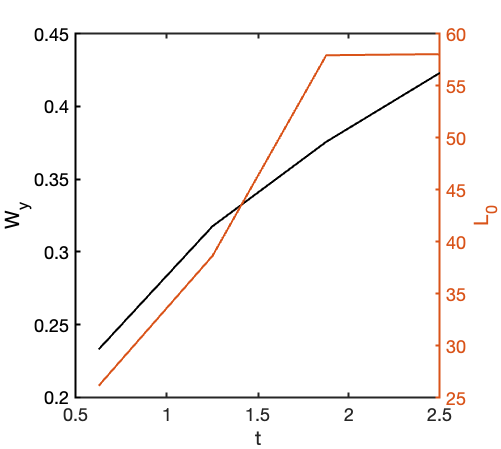}
    \caption{The skeleton length $L_0$ (orange) and the second moment of the skeleton $W_y$ (black) versus times, computed at the same times as in figure \ref{colorful2dskeletons}. }
    \label{moment}
\end{figure}

The total length of the skeleton $L_0$, its centroid $Y_0$ and its second moment $W_y$ are given by:
\begin{equation}
    L_0=\int ds; \qquad  Y_0=\frac{1}{L_0}\int y(s)ds; \qquad  W_y=\sqrt{\frac{1}{L_0}\int (y(s)-Y_0)^2ds},
    \label{momenteqn}
\end{equation}
where $y(s)$ is the $y$ location of each infinitesimal element on the skeleton and $ds$ is the length. Higher order moments can be defined in the same way. Notice that for a periodic jet, the centroid and the second moment in $x$ direction are meaningless. In figure \ref{moment} we see that the length of the skeleton increases with time, indicating an increase in elongated structures, since the total volume is conserved. The structures are also more concentrated near the jet axis at the initial time while later they spread out and become less compact, as shown by the second moment $W_y$ in figure \ref{moment}. 

To have a deeper look at the skeleton structure, we separate it into trunks and branches. Since the raw skeletons are actually thin films bounded by two curves (the original interface), we start by merging the curves into a single curve by setting a threshold distance $\epsilon$ and moving the points to the same coordinate if their distance is smaller than $\epsilon$ \citep{schirmacher1998boundary}. Notice that a carefully chosen $\epsilon$ may be necessary to get perfectly single-lined skeletons and avoid spurious branches. A large $\epsilon$ may eliminate important structures while a small $\epsilon$ is not able to merge curves into a single curve. Sometimes merging the curves with a small $\epsilon$ followed by merging the curves again with a larger $\epsilon$ will help to improve the quality. The choice of $\epsilon$ and number of times of merging may depend on problems. The skeletons processed in this way end up with several points having the same coordinates and we delete the redundant points. Then a network matrix can be constructed, which contains the number of points that a certain point is connected to. If a point has $2$ neighbours, then it is a normal connected point. Points that have 1 and 3 (or more than 3) neighbours are considered to be endpoints and junction points respectively. To identify branches, we start at an endpoint and move to the next connected point until we encounter a junction point or an endpoint. Then the counting stops and all the points on the path are considered to be a branch. This segmentation method is similar to the one used by \cite{palagyi2003quantitative} although their skeletons are defined on Eulerian grids. The definition of the hierarchy of branches follows the Horton-Strahler ordering (\cite{horton1945erosional,strahler1952hypsometric}), where a branch with an endpoint is order one level, the parent of two same-ordered branches is one order higher and the parent of two different ordered branches is assigned the same level as the branch with the higher level. The periodic ``trunk'' is assigned the highest level in the hierarchy. 

\begin{figure}
    \centering
    \includegraphics[scale=0.55]{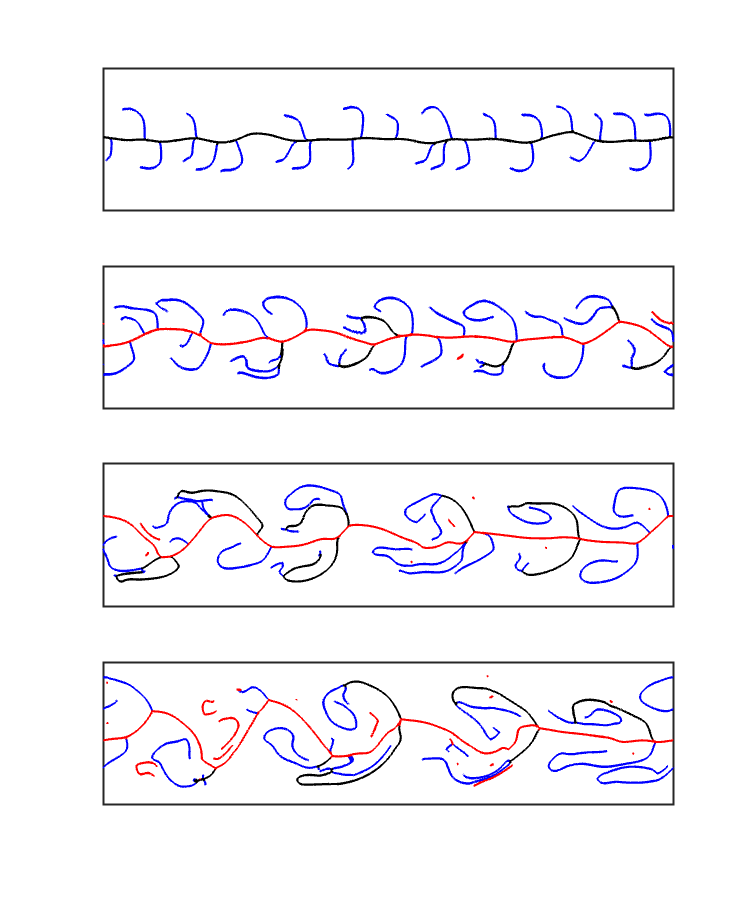}
    \caption{The skeleton of the jet after segmentation at 4 different times (corresponding to the times in figure \ref{colorful2dskeletons}). Different colors indicate different level: blue, black and red stand for first, second and third order level, respectively. Isolated branches are marked with the same color as the main trunk.}
    \label{segment}
\end{figure}

\begin{figure}[h]
    \centering
    {\includegraphics[scale=0.4]{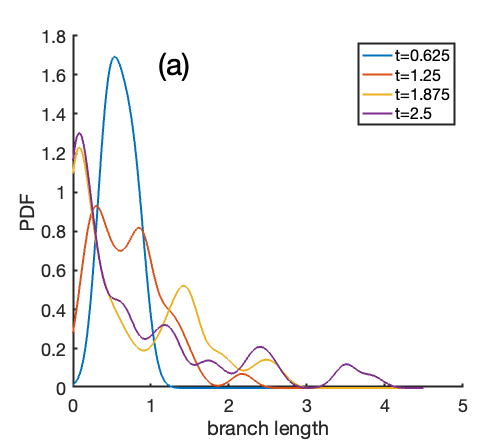} \includegraphics[scale=0.37]{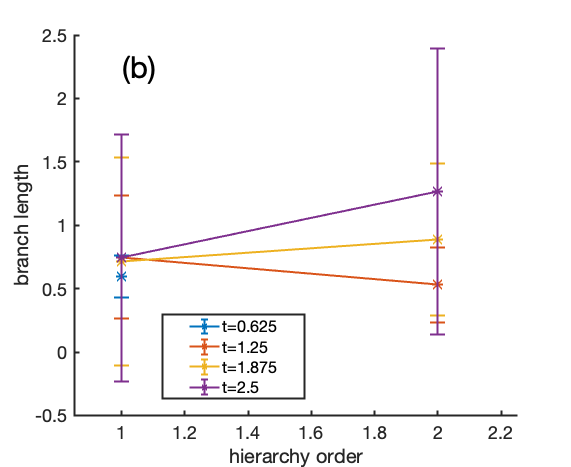} }
    \caption{(a) The pdf of the length of the branches for the four skeletons in figure \ref{segment}. (b) The average length of branches (marked by $*$) versus level for the four skeletons in figure \ref{segment}. The error bar indicates the standard deviation of the branch length within a certain level. }
    \label{bl}
\end{figure}

\begin{figure}
    \centering{\includegraphics[scale=0.5]{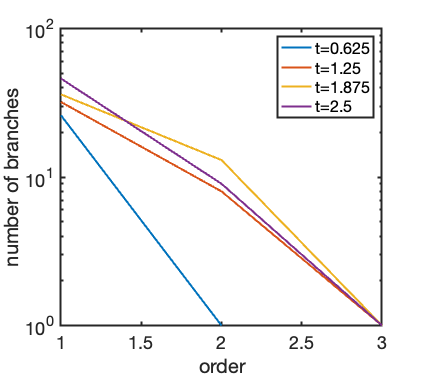} }
    \caption{The log of the number of branches versus level for the four skeletons in figure \ref{segment}.}
    \label{bifurcation}
\end{figure}

Figure \ref{segment} shows the skeletons after segmentation, using different colors to distinguish the different levels. At the initial stage, there are only two levels in the skeleton tree and the main trunk is the second order level. At the next time, some branches have their own ``children'' (branches) and the total number of levels increase to 3. In figure \ref{segment}(d), more isolated branches have been produced and they start to generate their own children. The branches length distribution and the average branch length versus different hierarchy order are plotted in figure \ref{bl}(a) and \ref{bl}(b), respectively. In figure \ref{bl}(a), it can be seen that at the initial stage all the branches are relatively short and uniform in length. But with time, the pdf distribution starts to skew toward the longer branches due to the growth of the branches. As some branches generate children branches, there is still a high proportion of short branches. At even later time the pdf is skewed more toward the longer branches but a larger number of short branches can also be seen, mainly due to an increasing number of short isolated branches. Figure \ref{bl}(b) shows that the standard deviation of the branch length is increasing with time for both the first and second order branches, showing that the branch length is becoming less uniform. Interestingly, the average lengths of the first order branches are about the same at the three later times, which may indicate that the breakup of the jet occurs when a critical branch length is reached. The average length of the second order branches, originally shorter than the first order branches, also keeps increasing in time and exceeds the average length of the first order branches. The number of branches versus the hierarchy order for four skeletons, in semi-log scale (\cite{brazeau1988inter,sanchez2003similar}), is plotted in figure \ref{bifurcation}. The number of first order branches increases with time while the number of second order branches first increases but then decreases due to breakups. We note that the bifurcation ratio $R_b$, defined as the ratio of the number of branches of a given order, to the number of branches of the next higher order, is a common quantitative measure in branching morphology, but here it is ill-posed due to the existence of the main periodic trunk. 

\subsection{Skeletons Equivalence}

How do we tell whether one skeleton is topologically different from another skeleton? The quantitative measures that we calculate above, such as number of branches, branch length and second order moment, characterize some aspects of the skeletons, but are not enough to determine the equivalence of the skeletons morphology. Equivalent obviously means a structure having the same quantitative measures, but the question of how much equivalence is needed may depend on the intended applications. 

\begin{figure}
    \centering
   { \includegraphics[width=0.6\textwidth]{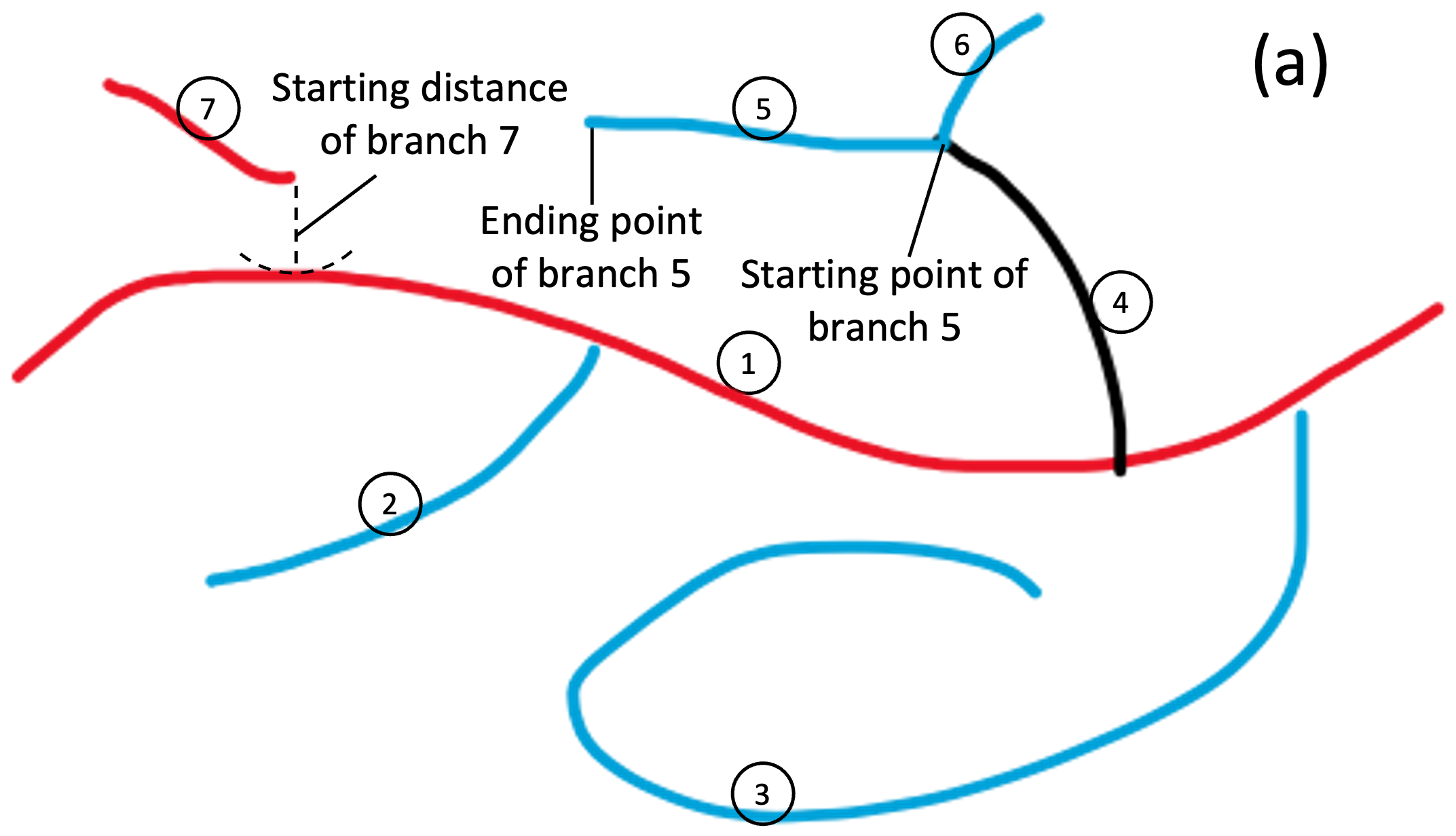} \\
    \includegraphics[width=0.45\textwidth]{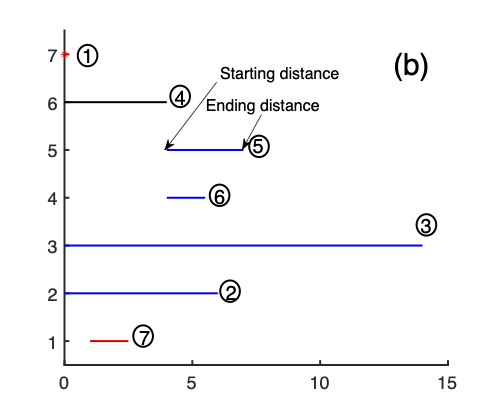}
    \includegraphics[width=0.45\textwidth]{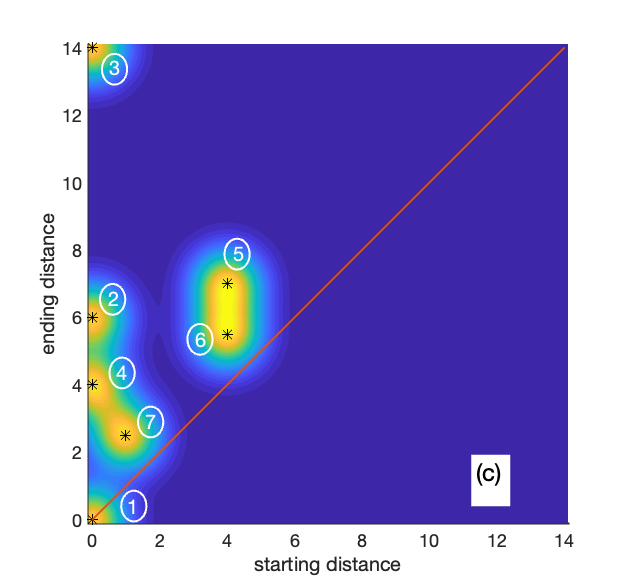}
    }
    \caption{(a) A schematic of a skeleton, where the numbers are the index for each branch and the colors correspond to their level. The colors are the same as the ones used in figure \ref{segment}. (b) The barcode for the skeleton. The numbers next to each line/point are the original indexes for each branch. (c) The persistence image of the skeleton, where a single $*$ stands for a certain branch and the original index of the branch is indicated next to the $*$. The $45^\circ$ line is shown in red. }
    \label{sketch}
\end{figure}

To analyze the structure of the skeleton in more detail, we use a slightly modified Topology Morphology Descriptor (TMD) proposed by \cite{kanari2018topological}. Note that we do not aim to reconstruct the skeletons fully from the extracted features, but seek a simplification that retains enough information to distinguish one skeleton structure from others. A schematic of a skeleton is shown in figure \ref{sketch}(a). First of all, we define the start and the end points of a branch as the points closest and furthest away from the trunk, along the skeleton path. The starting and ending distances are the path distance from the trunk to the start and the end points, respectively. The periodic trunk has $0$ for both starting and ending distance. If there was no periodic trunk, then the junction node of the highest level branch would be used as a reference point (or ``root'') for calculating the path distance. The starting distance of an isolated branch (branch $7$ in figure \ref{sketch}(a)) is defined as the shortest point-to-point distance of the endpoint in the branch to the main trunk. The ending distance is simply the length of the branch plus the starting distance. After defining the starting and the ending distance for all the branches, we sort the trunk and the isolated branches in a descending order (branch $1,7$), based on the total length of the skeleton tree, defined as the combined structure with all interconnecting branches (branch $1-6$). Then we find the children of each branch and sort them first in terms of the level and then by the length in descending order (for sorted children of the trunk $1$: branch $4,3,2$). If the branches have their own children, they are sorted in the same way. Finally, we can draw horizontal lines for each branch, where the left and right endpoints indicate the starting and ending distance, in the sorted order. This is referred to as a barcode by \citet{kanari2018topological}. Figure \ref{sketch}(b) shows the barcode for the skeleton in \ref{sketch}(a), where the vertical axis is the index of the branch after being sorted and $x$ axis is the distance. The barcode is intended to make the whole structure of the skeleton tree, and the relationship between each branch, as clear as possible. The branches are colored according to their level so that the branches that are connected to the trunk and the isolated branches can be easily distinguished. The barcode is also independent of the coordinate system being used. In principle, we can use point-to-point distance rather than the path distance to define the starting and ending, as in \cite{kanari2018topological}, but we found that the current definition is more suitable for this problem, especially for the case that includes a periodic trunk and isolated branches. 

We can further compress the information by representing each branch as a point in a plane where the $x$ axis is the starting distance and the $y$ axis is the ending distance, as shown in figure \ref{sketch}(c). \citet{kanari2018topological} refer to this as the persistence image. To get a smooth distribution, we apply a Gaussian kernel $G=e^{\frac{-|\mathbf{x}-\mathbf{x'}|^2}{\gamma^2}}$ to those points with $\gamma=1.0$ and then normalize so that the integral of the persistence image over all the domain is $1$. Notice that unlike the persistence image used in \cite{kanari2018topological,kanari2019objective}, where there can be points both above and below the $45^{\circ}$ line, since the point-to-point distance is used rather than path distance, in our persistence image there can only be points above the $45^{\circ}$ line.

\begin{figure}
    \centering
    {\includegraphics[scale=0.33]{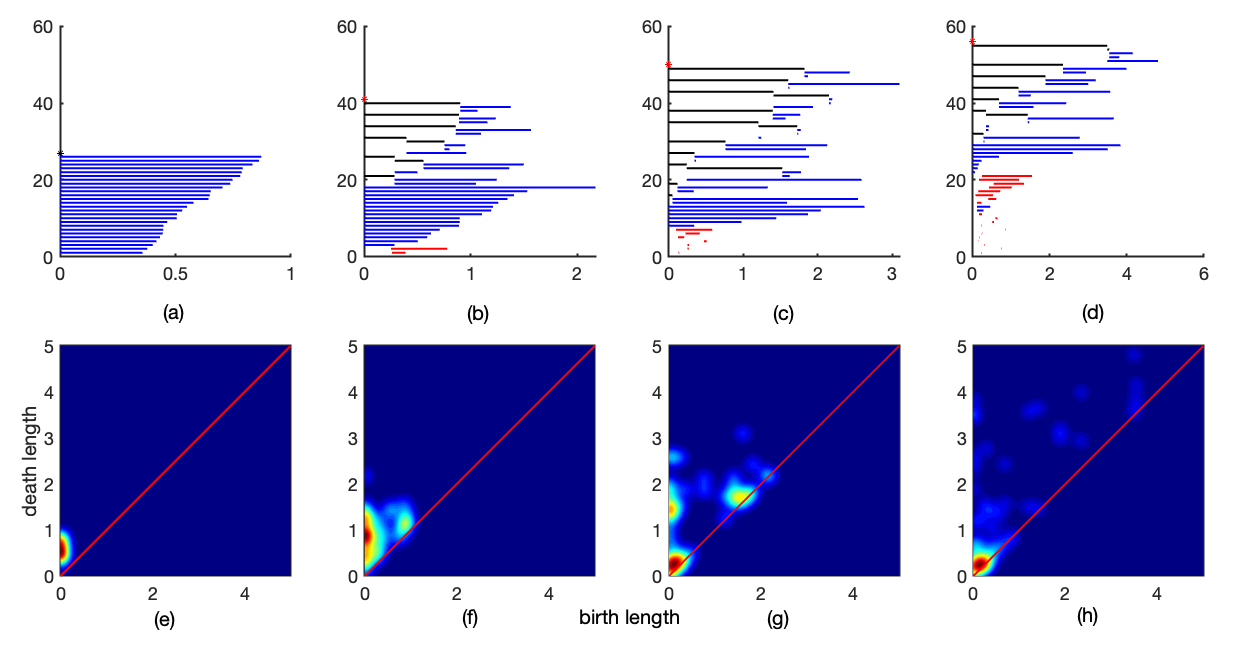}  }
    \caption{The barcodes colored based on the hierarchy of each branch (top row) and the persistence images (bottom row) for each of the 4 skeletons, from left to right corresponding to the skeletons in figure \ref{segment} from top to bottom. $45^{\circ}$ lines are drawn in red in the persistence images. Here we use $\gamma=0.2$ to smooth the persistence images.}
    \label{tmd}
\end{figure}

Figure \ref{tmd} shows the barcodes ((a)-(d)) and the persistence images ((e)-(h)) of the skeletons in figure \ref{segment}. It is straightforward to tell from the barcodes that more branches, children and isolated branches are generated with time. At the initial stage (figure \ref{tmd}(e)), the persistence image shows a high concentration around the bottom left corner on the $y$ axis, since all the branches are short and connected directly to the trunk. As time increases (figure \ref{tmd}(f)), the patch in the bottom left corner is elongated in the $y$ direction and a separate patch occurs on the right, because branches start to generate children and their starting distances are not $0$. Later, branches grow in length and generate more children, resulting in the patches being spread out (figures \ref{tmd}(g) and (h)). Notice that points that sit close to the $45^{\circ}$ line indicate short branches. The high concentration around the origin in those two figures reflects a significant number of short branches floating around the main trunk. In contrast to the initial compact concentration along the $y$ axis, the persistence image at the later time shows a more expanded and uniform distribution, except for a high concentration around the origin. 

\begin{figure}
    \centering {
    \includegraphics[width=0.55\textwidth]{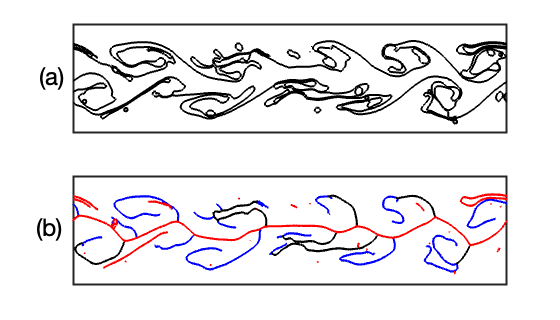}
    \includegraphics[width=0.4\textwidth]{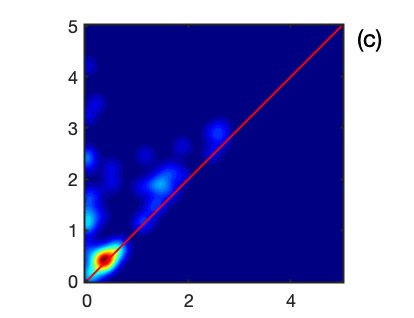} }
    \caption{(a) The original interface at $t=2.5$. (b) The corresponding skeleton colored based on the hierarchy level, in the same way as in figure \ref{segment}. (c) The corresponding persistence image with $45^{\circ}$ line plotted in red. }
    \label{persistence_comparison}
\end{figure}

We believe that the persistence image is a good way to characterize the skeleton structures and can be used to evaluate the similarities and differences between different structures. To explore this we have simulated the evolution of a jet with exactly the same flow parameters as in figure \ref{colorful2dskeletons} but with different initial perturbations, and show the interface in figure \ref{persistence_comparison}(a) at $t = 2.5$. The shape is very different from the one in figure \ref{colorful2dskeletons}(g) but to evaluate whether they share similar characteristics we extract the skeleton (figure \ref{persistence_comparison}(b)) and plot the persistence image in figure \ref{persistence_comparison}(c). While it is not identical to the persistence images in figure \ref{tmd}(h) there are considerable similarities and we can conclude that those two jets are topologically similar. \cite{kanari2018topological} and \cite{kanari2019objective} used persistence images to do classification of neuronal trees, but here we have not attempted to do that. If the main jet breaks, so the main trunk does not exist, the persistence image can presumably still distinguish the structures. For example, if the skeleton in figure \ref{segment}(d) stays unchanged except for the trunk breaking in the middle, the distribution in the persistence image will be more uniformly distributed and not include a high concentration around the origin, since the starting distance from those isolated branches to the root is increased. 

\subsection{Skeletonization in 3D}

The extension of the skeletonization algorithm from 2D to 3D interfaces is straightforward and here we show one example. The computational setup is the same as used in \citet{afanador2021effect}, giving the nondimensional numbers $Re=\rho_j \Delta u d/\mu_j=150$ and $We=\rho_j\Delta u^2d/\sigma=300$. We used a $256\times128\times128$ gird to solve for the flow, but in order to improve the quality of the skeleton, we took the instantaneous interface topology and constructed the indicator function on a finer grid ($512\times256\times256$) for the skeletonization process.

\begin{figure}
    
    \centering{
    \includegraphics[scale=0.3]{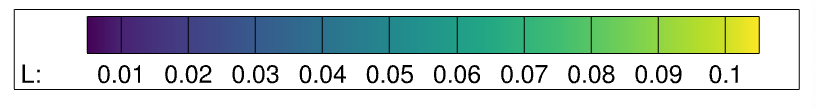}
    \includegraphics[scale=0.19]{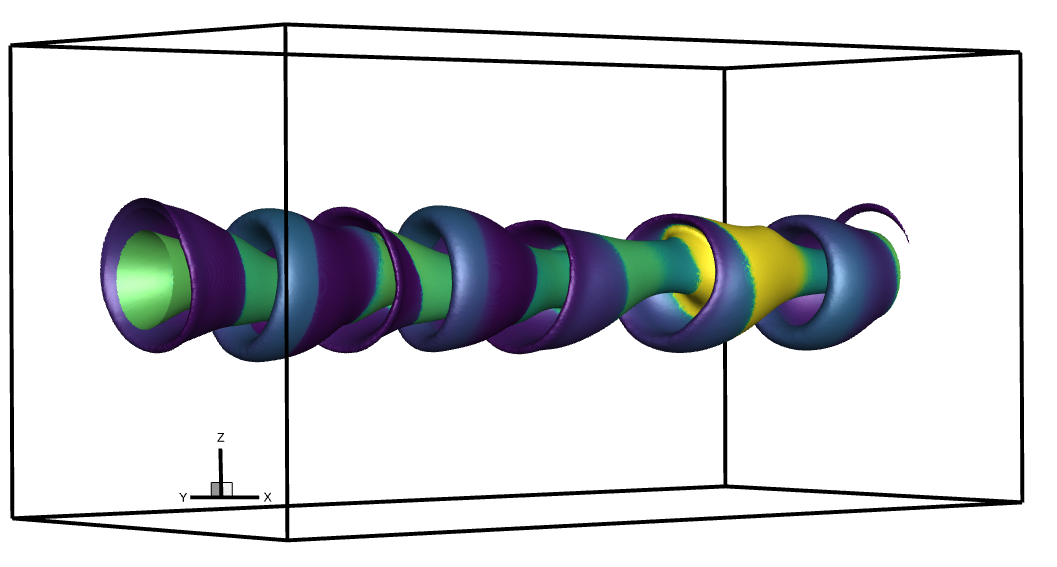} \includegraphics[scale=0.19]{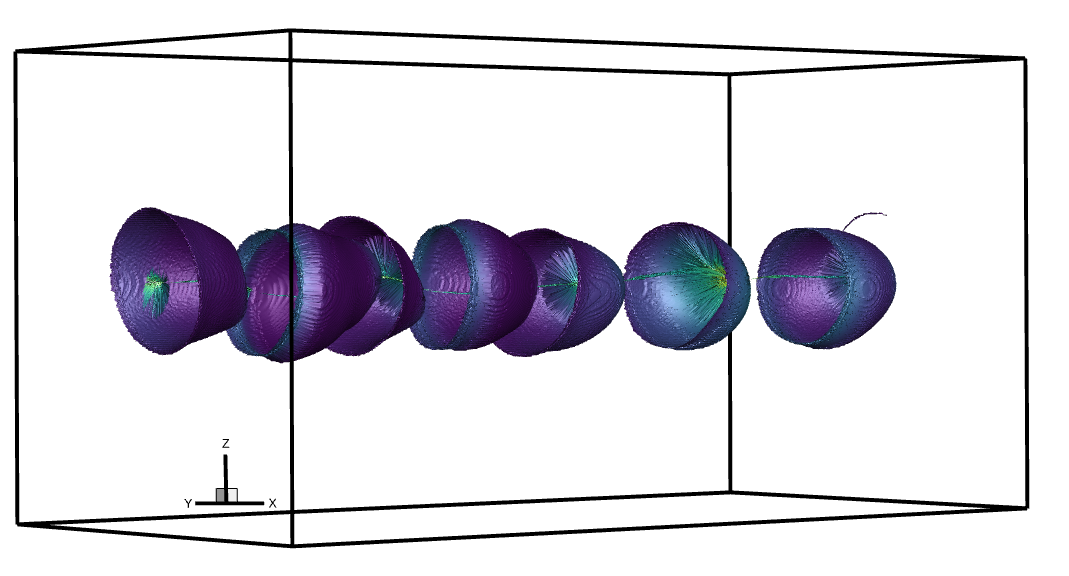} \includegraphics[scale=0.19]{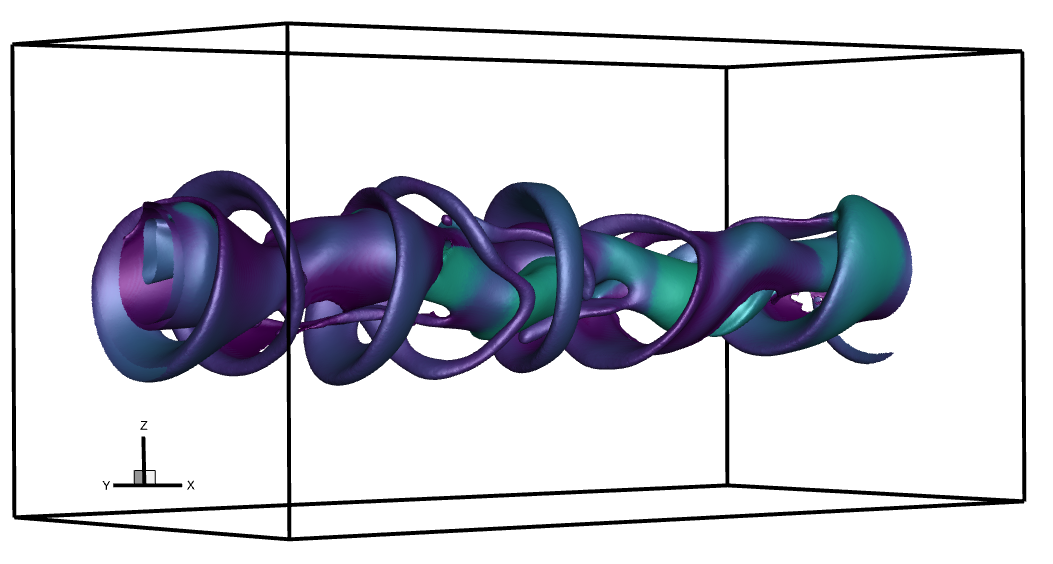} \includegraphics[scale=0.19]{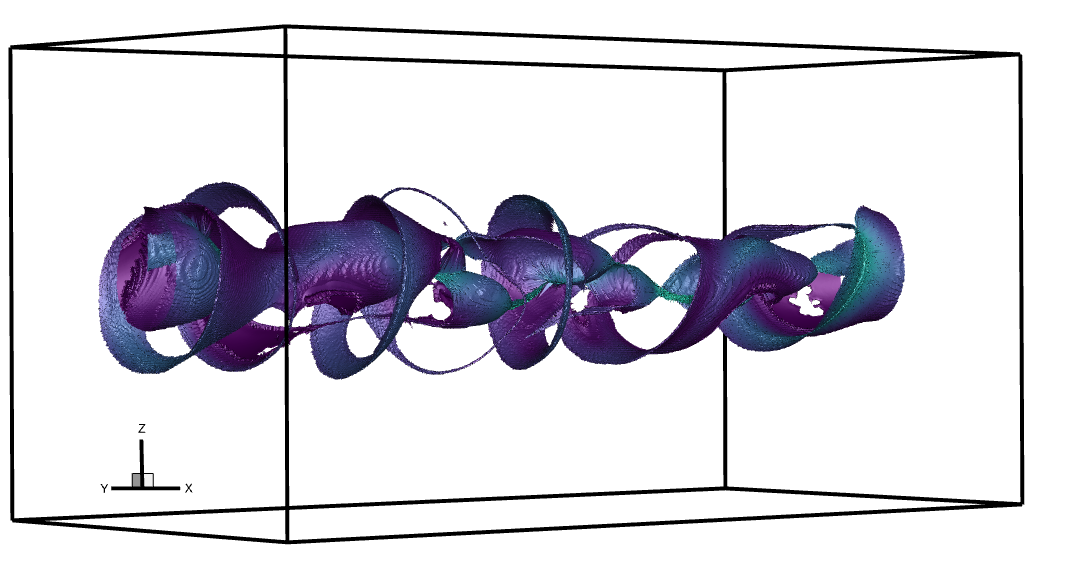}
    \includegraphics[scale=0.19]{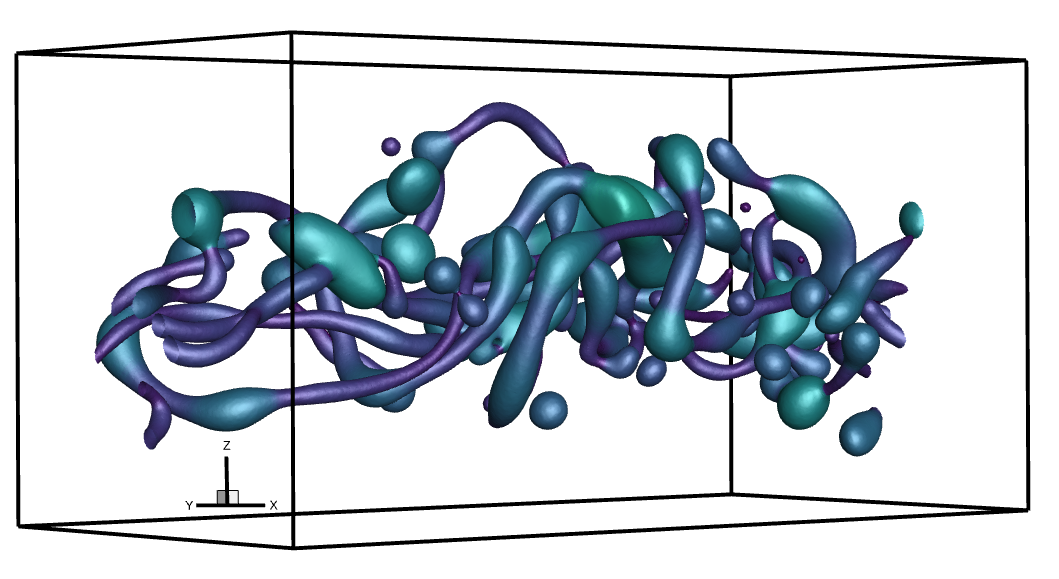}
    \includegraphics[scale=0.19]{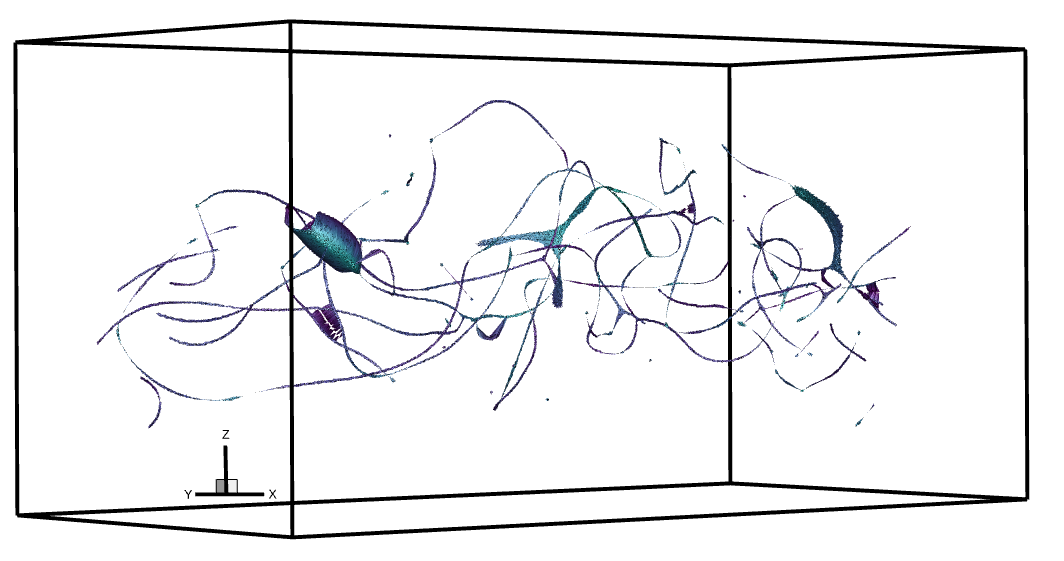}
    
    }
    \caption{On the left plots the original interface, colored based on local length scale $L_s$. The skeletons are plotted on the right, colored based on the corresponding length scale $L$. From top to bottom the corresponding computational time is $t_1=2.5$, $t_2=5$ and $t_3=15$.}
    \label{3dskeleton}
\end{figure}

Figure \ref{3dskeleton} shows the three dimensional jet at three times (left) and the corresponding skeleton (right). At the two early times, parts of the jet have collapsed into 2D thin sheets, rather than 1D filaments, shown in the first two rows. Those are obviously not seen for 2D flows. The sheets are similar to a medial surface defined as the union of the center points of inscribed balls \citep{cornea2007curve}, although the skeletons generated by our algorithm are not guaranteed to be exactly centered. The formation of the 2D sheets indicates a high aspect ratio interface topology, where the length scale in one direction is much smaller than in the other two directions. When the jet is about to break into ligaments we see 1D filaments, corresponding to the 2D case, as shown in the second row. The formation of a 1D skeleton indicates a structure where the dimensions in two directions are about the same and much smaller than in the third direction. The last row shows a jet that has completely disintegrated and the skeleton consists mostly of strings and points, which are topologically different from what we see in the first 2 rows. The skeletons show that the flow field involves mostly long filaments and spherical drops. A small number of surface skeletons can also be found at the last time (bottom row), bringing out an aspect of the topology of the interface which is not obvious from the original interface. Thus, the skeleton representation is shown to be useful in visualizing and understanding the topology of the jet. Segmenting the skeletons into different parts for this case is challenging due to the existence of the surface skeletons and new quantitative measures are required to distinguish surface skeletons from line skeletons. Using the area tensor (\cite{wetzel1999area}) to find the projected area on the principal directions may be one possibility. TMD, including the barcodes and the persistence images, would need to be redesigned to characterize the skeleton topology to account for the surface skeletons. 

\section{Conclusion}
We examine the use of skeletonization to visualize and characterize a liquid structure in multiphase flows in both two and three-dimensions, by representing thin ligaments with 1D-lines (or 2D-sheets) and spherical drops as points. A way to systematically skeletonize a fluid strucure is proposed, by gradually shrinking the surface using diffusion of an index function, until it has collapsed into thin skeletons. The skeletons are generated in a robust way that ignores boundary noise, retains one-to-one correspondence to the original interface, and is able to reveal the underlying basic topology of complicated fluid structures, as well as allowing us to approximately reconstruct the original structures. Quantitative measures for the skeletons for 2D flows, including branch length distribution and second order moment, provides useful information about the topology of the jet. A more sophisticated ``equivalent'' compact description, TMD, compresses the information further and generates persistence images so that different skeletons can be distinguished topologically.

We expect that this work can be extended to build reduced order models for multiphase flows, by compressing one phase into skeletons and evolving the skeletons with a filtered flow field. Representing the complicated topology with skeletons in a systematic way, along with quantitative measures of skeletons are also potentially useful in studying the underlying topology of different structures, including interfaces in multiphase flows and vortex cores.

\section*{Acknowledgement}
This research was supported in part by the National Science Foundation Grant CBET-1953082 and by ERC Advanced Grant TRUFLOW. Computations were done at the Advanced Research Computing at Hopkins (ARCH) core facility (rockfish.jhu.edu), which is supported by the National Science Foundation (NSF) grant number OAC 1920103.

\appendix
\appendixpage
\section{Controlling the sensitivity of the skeletonization}
\label{appendix}
\begin{figure}
    \centering{\includegraphics[scale=0.4]{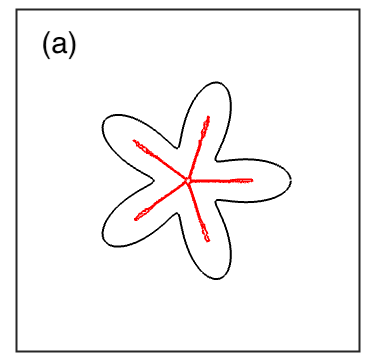} \includegraphics[scale=0.4]{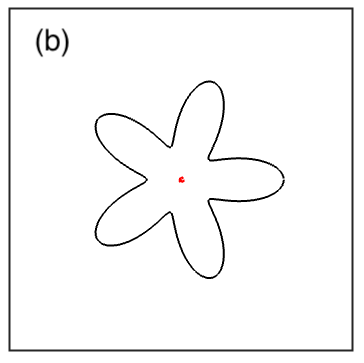} }
    \caption{The original interface (black) and its skeleton (red), with (a) constant diffusion coefficient $D$ and (b) nonuniform diffusion coefficient that depends on curvature $D(\kappa)=1+8|\kappa|/\max(|\kappa|)$.}
    \label{star}
\end{figure}

In order to control the sensitivity of the skeletonization to interface noise, we can make the diffusion coefficient $D$ dependent on the local curvature $\kappa$ and solve a nonlinear diffusion equation, based on
\begin{equation}
\frac{\partial \chi }{ \partial \tau} =  \nabla\cdot D(\kappa)\nabla \chi,
\label{diffusion1}
\end{equation}
to move the interface with a specific contour. Figure \ref{star} shows a comparison of the skeletonization with a constant diffusion coefficient and a curvature dependent one. By taking the diffusion coefficient to be large in high curvature regions, the noise is quickly eliminated and the skeleton is less sensitive to the boundary shape. Here, we use $D(\kappa)=1+8|\kappa|/\max(|\kappa|)$. We note that this is just one way to change the diffusion coefficient and we have made no effort to optimize $D(\kappa)$, in any way. 

\bibliography{tc}

\end{document}